\documentclass[aps,prl,preprintnumbers,amsmath,amssymb,latexsym,nofootinbib,array,enumerate,letter,twocolumn,superscriptaddress]{revtex4-1}

\usepackage{here}
\usepackage{comment,braket}
\usepackage{epsf}
\usepackage{amsmath}
\usepackage{graphics}
\usepackage{amsfonts}
\usepackage{amssymb}
\usepackage{latexsym}
\usepackage{color}
\usepackage{ulem}
\input{colordvi.tex}
\usepackage{feynmp}

\usepackage[pdftex]{graphicx}
\usepackage{bm}
\usepackage[pdftex]{hyperref}
\usepackage[svgnames]{xcolor}
\definecolor{phthaloblue}{rgb}{0.0, 0.06, 0.54}
\hypersetup{
    colorlinks=true,
    linkcolor=phthaloblue,
    citecolor=blue,
    filecolor=blue,
    urlcolor=phthaloblue,
    }

\newcommand{\beq}{\begin{eqnarray}} 
\newcommand{\eeq}{\end{eqnarray}}

\def\({\left(}
\def\){\right)}

\newcommand{\be}{\begin{equation}}
\newcommand{\ee}{\end{equation}}
\newcommand{\bea}{\begin{array}} 
\newcommand{\eea}{\end{array}}

\newcommand{\GeV}{\  {\rm GeV} }

\begin{document}

\title{
Axion Kinetic Misalignment Mechanism
}
\preprint{LCTP-19-28}

\author{Raymond T. Co}
\affiliation{Leinweber Center for Theoretical Physics, University of Michigan, Ann Arbor, MI 48109, USA}
\author{Lawrence J. Hall}
\affiliation{Department of Physics, University of California, Berkeley, California 94720, USA}
\affiliation{Theoretical Physics Group, Lawrence Berkeley National Laboratory, Berkeley, California 94720, USA}
\author{Keisuke Harigaya}
\affiliation{School of Natural Sciences, Institute for Advanced Study, Princeton, NJ 08540, USA}

\begin{abstract}
In the conventional misalignment mechanism, the axion field has a constant initial field value in the early universe and later begins to oscillate. We present an alternative scenario where the axion field has a nonzero initial velocity, allowing an axion decay constant much below the conventional prediction from axion dark matter. This axion velocity can be generated from explicit breaking of the axion shift symmetry in the early universe, which may occur as this symmetry is approximate.
\end{abstract}

\date{\today}

\maketitle

{\bf Introduction.}---%
Why is CP violation so suppressed in the strong interaction~\cite{tHooft:1976rip, Crewther:1979pi, Baker:2006ts} while near maximal in the weak interaction?  The Peccei-Quinn (PQ) mechanism~\cite{Peccei:1977hh,Peccei:1977ur} provides a simple and elegant answer: the angular parameter describing CP violation in the strong interaction is actually a field resulting from spontaneous symmetry breaking, $\theta(x)$. A potential $V(\theta)$ arises from the strong interaction and has CP conserving minima, as shown in Fig.~\ref{fig:cartoons}.  Axions are fluctuations in this field \cite{Weinberg:1977ma,Wilczek:1977pj} and the mass of the axion is powerfully constrained by particle and astrophysics, $m_a < 60$ meV; equivalently, there is a lower bound on the PQ symmetry breaking scale  $f_a = 10^8 \; \mbox{GeV} \left(60 \; \mbox{meV} / m_a \right)$~\cite{Ellis:1987pk,Raffelt:1987yt,Turner:1987by,Mayle:1987as,Raffelt:2006cw,Chang:2018rso,Carenza:2019pxu}.

In the early universe, if the initial value of the field, $\theta_i$, is away from the minima, the axion field starts to oscillate at a temperature $T_*$ when $m_a \sim 3H$, where $H$ is the Hubble expansion rate.  These oscillations, illustrated in the upper diagram of Fig.~\ref{fig:cartoons}, can account for the observed dark matter \cite{Preskill:1982cy,Abbott:1982af,Dine:1982ah}.
For $\theta_i$ not accidentally close to the bottom nor the hilltop of the potential, 
this ``misalignment" mechanism predicts an axion mass of order $10~\mu$eV and tends to underproduce for heavier masses.

In this Letter we show that an alternative initial condition for the axion field, $\dot{\theta} \neq 0$, leads to axion dark matter for larger values of $m_a$. This ``kinetic misalignment" mechanism is operative if the axion kinetic energy is larger than the potential energy at temperature $T_*$, delaying the onset of axion field oscillations, as shown in the lower diagram of Fig.~\ref{fig:cartoons}.  We begin with an elaboration of the basic mechanism. We then show that a sufficient $\dot{\theta}$ can arise at early times from explicit breaking of the PQ symmetry by a higher dimensional operator in the same manner as the Affleck-Dine mechanism, which generates rotations of complex scalar fields~\cite{Affleck:1984fy,Dine:1995kz}.

The PQ symmetry is an approximate symmetry which is explicitly broken by the strong interaction. It is plausible that higher dimensional operators also explicitly break the PQ symmetry. Although they should be negligible in the vacuum in order not to shift the axion minimum from the CP conserving one, they can be effective in the early universe if the PQ symmetry breaking field takes a large initial value. Higher dimensional PQ-breaking operators are in fact expected if one tries to understand the PQ symmetry as an accidental symmetry arising from some exact symmetries~\cite{Holman:1992us,Barr:1992qq,Kamionkowski:1992mf,Dine:1992vx}. The kinetic misalignment mechanism is therefore a phenomenological prediction intrinsically tied to the theoretical origin of the PQ symmetry.

\begin{figure}[t]
\begin{center}
\includegraphics[width=\linewidth]{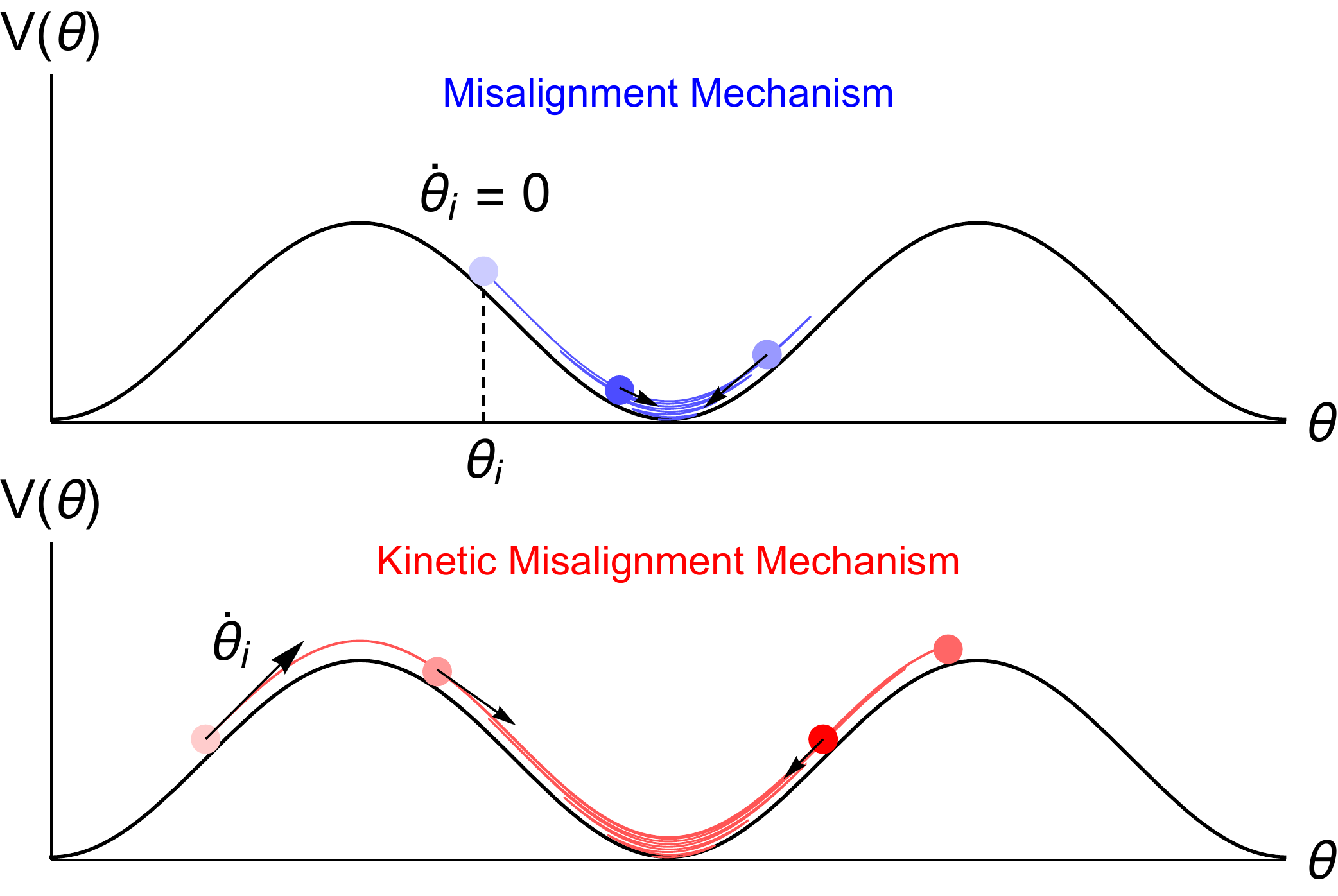}
\caption{
The schematics of the (kinetic) misalignment mechanism. Initial conditions are labeled, shadings from light to dark indicate the time sequence of the motion, and arrows with different relative lengths denote instantaneous velocities.
}
\label{fig:cartoons}
\end{center}
\end{figure}

The mechanism allows for axion dark matter with a mass above the prediction of the standard misalignment mechanism. This mass scale $m_a = \mathcal{O}(0.1$-$100)$ meV is under extensive experimental investigation~\cite{Vogel:2013bta, Armengaud:2014gea, Arvanitaki:2014dfa, Rybka:2014cya, Sikivie:2014lha, TheMADMAXWorkingGroup:2016hpc, McAllister:2017lkb, Anastassopoulos:2017kag, Arvanitaki:2017nhi, Geraci:2017bmq, Baryakhtar:2018doz,Du:2018uak, Marsh:2018dlj, Lawson:2019brd, Zarei:2019sva}.
Other known production mechanisms in this mass range are 1) parametric resonance from a PQ symmetry breaking field~\cite{Co:2017mop, Harigaya:2019qnl}, 2) anharmonicity effects~\cite{Turner:1985si,Lyth:1991ub,Visinelli:2009zm} when $\theta_{i}$ approaches $\pi$ due to fine-tuning or inflationary dynamics~\cite{Co:2018mho, Takahashi:2019pqf}, 3) decays of unstable domain walls \cite{Sikivie:1982qv,Chang:1998tb,Hiramatsu:2010yn,Hiramatsu:2012sc,Kawasaki:2014sqa,Ringwald:2015dsf,Harigaya:2018ooc,Caputo:2019wsd}, and 4) production during a kination era~\cite{Visinelli:2009kt}.
Contrary to these mechanisms, kinetic misalignment offers an exciting theoretical connection with the baryon asymmetry of the Universe through so-called axiogenesis~\cite{Co:2019wyp}.

{\bf Kinetic misalignment mechanism.}---%
We estimate the dark matter abundance for a generic axion-like field with decay constant $f_\phi$, $\phi = f_\phi \theta$,  when $\dot{\theta} \neq 0$. Without loss of generality we take $\dot{\theta} >0$.   It is convenient to express the rotation by the dimensionless quantity $\dot\theta f_\phi^2/s$
where $s \propto R^{-3}$ is the entropy density and $R$ is the scale factor. In fact, $\dot\theta f_\phi^2$ is the Noether charge associated with the shift symmetry $\phi \rightarrow \phi + \alpha f_\phi $ and hence should decrease in proportion to $R^{-3}$. Therefore, $\dot\theta f_\phi^2/s$ remains constant.
We define
\begin{equation}
\label{eq:n_Y_theta}
n_\theta \equiv \dot\theta f_\phi^2, \hspace{0.5 in} Y_\theta \equiv \frac{n_\theta}{s} ,
\end{equation}
and call $n_\theta$ a charge density and $Y_\theta$ a yield. Even though $Y_\theta$ is a redshift-invariant quantity, the evaluation of $Y_\theta$ from the initial condition is model-dependent and is thoroughly discussed later in this Letter. 

We will eventually consider the case where the axion originates from a phase direction of a complex scalar field $P$ whose vacuum expectation value spontaneously breaks a $U(1)$ global symmetry,
\begin{equation}
P \equiv \frac{1}{\sqrt{2}} \left( S + f_\phi \right) e^{ i \frac{\phi}{f_\phi} },
\end{equation}
where $S$ and $\phi$ are the radial and angular (axion) modes respectively. However, the kinetic misalignment mechanism can be understood without referring to $P$.
The nonzero axion velocity corresponds to a rotation of $P$.

We assume the potential of the axion is
\begin{equation}
V = m_\phi(T)^2 f_\phi^2 \left(1 - {\rm cos}\frac{\phi}{f_\phi} \right),
\end{equation}
where the axion mass $m_\phi(T)$ may depend on temperature $T$.
If $Y_\theta$ is sufficiently small, axion field oscillations begin at $T_*$ where $m_\phi(T_*) = 3H(T_*)$, yielding the conventional misalignment mechanism.

Our key point is that, if the axion kinetic energy $K = \dot\theta^2 f_\phi^2/2$ is larger than its potential energy $V(\phi)$ at the conventional oscillation temperature $T_*$, the axion simply overcomes the potential barrier and the misalignment angle continues to change at the rate $\dot\theta (T)$. This evolution ceases when the kinetic energy $K$ redshifts to the height of the potential barrier, $V_{\rm max} = 2 m_\phi^2(T) f_\phi^2$, at the temperature we call $T'$,
\begin{equation}
\label{eq:dtheta_mphi}
\dot\theta (T') = 2 m_\phi(T'). 
\end{equation}
Subsequently, the axion is trapped by the potential barrier and oscillates around the minimum. The onset of oscillation is delayed if this trapping happens after the conventional oscillation temperature, $T' < T_*$. Equivalently, kinetic misalignment is at play when 
\begin{equation}
\label{eq:mphi_Hubble}
m_\phi(T') \geq 3 H(T') .
\end{equation}

If the axion mass changes adiabatically, the number density is conserved. The energy density of the axion oscillation $\rho_\phi$ normalized by the entropy density reads
\begin{align}
\frac{\rho_\phi}{s} & = m_\phi(0) \frac{n_\phi (T')}{s(T')} = C m_\phi(0) Y_\theta .
\label{eq:rho_s_dtheta}
\end{align}
The axion abundance only depends on $Y_\theta$ and the mass of the axion in the vacuum, and is independent of the evolution of the axion mass. 

The analytic estimate thus far predicts $C = 1$ since 
\begin{equation}
n_\phi (T') = \frac{V_\phi }{ m_\phi(T') } \simeq  2m_\phi(T')f_\phi^2  \simeq \dot\theta(T') f_\phi^2 = n_\theta .
\end{equation}
However, this estimate is not precise because $\rho_\phi$ is assumed to scale as $R^{-3}$ as soon as the oscillation starts at $T'$. Since the oscillation starts near the top of the cosine potential where the potential gradient is small, there is a further delay in the oscillation and this anharmonicity enhances the axion abundance. This nonlinear effect calls for a numerical analysis, which we perform in the Supplemental Material and determine $C \simeq 2$.

The kinetic misalignment mechanism is effective when
\begin{equation}
\label{eq:Y_theta_crit}
Y_\theta >   Y_{\rm crit} = \frac{n_\phi(T_*)}{s(T_*)} \sim \frac{f_\phi^2}{M_{\rm Pl} T_*} .
\end{equation}
Furthermore, sufficient axion dark matter results if $m_\phi(0) Y_\theta \simeq T_e$, where $T_e$ is the temperature of matter-radiation equality. The condition for kinetic misalignment in Eq.~(\ref{eq:Y_theta_crit}) then requires $m_\phi(0) f_\phi^2 \lesssim T_e T_* M_{\rm Pl}$.

We now estimate $Y_{\rm crit}$ for the QCD axion, $a$, and take  $m_a \propto T^{-4}$ for $T> \Lambda_{\rm QCD}$ from the dilute instanton gas approximation (also see the lattice results in~\cite{Petreczky:2016vrs, Borsanyi:2016ksw, Burger:2018fvb, Bonati:2018blm, Gorghetto:2018ocs}),
\begin{equation}
\label{eq:YP_crit}
 Y_{\rm crit} = 0.11  \left( \frac{f_a}{10^9~{\rm GeV}} \right)^{ \scalebox{1.01}{$\frac{13}{6}$} }.
\end{equation}
For $Y_\theta \gg Y_{\rm crit}$ the axion abundance is
\begin{equation}
\label{eq:Yv_axion}
\Omega_a h^2 \simeq \Omega_{\rm DM} h^2 \left( \frac{10^9~{\rm GeV}}{f_a} \right) \left( \frac{Y_\theta}{40} \right),
\end{equation}
which is independent of the axion mass evolution. For $f_a \gtrsim 1.5 \times 10^{11} \GeV$, kinetic misalignment cannot yield axion dark matter, since Eqs.~(\ref{eq:YP_crit}) and (\ref{eq:Yv_axion}) then give $Y_\theta < Y_{\rm crit}$ and the usual misalignment mechanism results.

The relevant numerical results and analytic derivations are thoroughly presented in the Supplemental Material.

{\bf Rotation from higher dimensional operators.}---%
Assuming that the potential of $|P|$ is sufficiently flat, a large field value may arise during inflation as an initial condition, by quantum fluctuations, or due to a negative Hubble-induced mass. For large enough initial field value $|P_i|$, the explicit breaking of the global symmetry by higher dimensional operators may become important. Such operators give a potential gradient to the angular direction of $P$ and drive angular motion. By the cosmic expansion, the field value $|P|$ decreases and the higher dimensional operator becomes ineffective. The angular direction then has a flat potential and $P$ rotates about the origin. This dynamics is the same as that in Affleck-Dine baryogenesis~\cite{Affleck:1984fy,Dine:1995kz} with supersymmetric partners of quarks and leptons.

The rotation is understood as a state with an asymmetry of the global charge. The density of the Noether charge associated with the symmetry $P \rightarrow e^{i \alpha}P$ is
\begin{equation}
n_\theta = i P \dot{P}^* -i P^* \dot{P},
\end{equation}
which is nonzero for a rotating $P$ and reduces to Eq.~(\ref{eq:n_Y_theta}) when $|P|$ is relaxed to $f_\phi/\sqrt{2}$. 

At the onset of the rotation, the asymmetry is
\begin{equation}
\label{eq:n_theta}
n_\theta = \epsilon \frac{V(P_i)}{m_S (P_i)}, \hspace{0.5 in} \epsilon \lesssim 1,
\end{equation}
where $\epsilon$, defined by this equation, parametrizes how close the trajectory is to a circular motion, which maximizes the asymmetry for a fixed energy. The size of $\epsilon$ is determined by the potential gradient of the angular direction relative to that of the radial direction.

Soon after the onset of the rotation, $n_\theta R^3$ becomes a conserved quantity, implying that $\dot\theta \propto R^{-3} |P|^{-2}$ redshifts slower for $|P| \gg f_a$ than if $|P|$ is fixed at $f_\phi / \sqrt{2}$. This slower redshift plays an important role and may explain why a $\dot\theta$ sufficiently large to affect the axion dynamics around the QCD phase transition has been neglected. In what follows, we explicitly demonstrate the kinetic misalignment mechanism using a quartic and a quadratic potential.

{\bf Model with Quartic Potential.}---%
We first demonstrate kinetic misalignment with the quartic potential for the global symmetry breaking field $P$,
\begin{equation}
\label{eq:quartic}
V =  \lambda^2 \left( |P|^2 - \frac{f_\phi^2}{2} \right)^2, \hspace{0.3 in} \lambda^2 = \frac{1}{2} \frac{m_S^2}{f_\phi^2},
\end{equation}
where $m_S$ is the vacuum mass of the radial degree of freedom $S$. Simply following the terminology in supersymmetric theories, we call $S$ the saxion. For small $\lambda$, the saxion has a flat potential and may obtain a large field value during inflation. At an initial field value $|P_i| = S_i/ \sqrt{2}$, the saxion mass is $\sqrt{3} \lambda S_i$. The saxion begins to oscillate when the mass exceeds $3H$. Assuming radiation domination, the temperature at which this occurs is
\begin{equation}
\label{eq:Tosc_Si}
T_{\rm osc} \simeq 2 \times 10^{12}~{\rm GeV} \left( \frac{S_i}{10^{17}~{\rm GeV}} \right)^{ \scalebox{1.01}{$\frac{1}{2}$} }  \left( \frac{\lambda }{10^{-10}} \right)^{ \scalebox{1.01}{$\frac{1}{2}$} }.
\end{equation}

When the oscillation starts, the asymmetry given in Eq.~(\ref{eq:n_theta}) is $n_\theta = \epsilon \lambda S_i^3 / (4 \sqrt{3})$, corresponding to the yield
\begin{equation}
\label{eq:YP4}
Y_\theta \equiv \frac{n_\theta}{s} \simeq 40 \  \epsilon \left( \frac{S_i}{10^{17}~{\rm GeV}} \right)^{ \scalebox{1.01}{$\frac{3}{2}$} } \left( \frac{10^{-10}} {\lambda}\right)^{ \scalebox{1.01}{$\frac{1}{2}$} },
\end{equation}
which remains constant unless entropy is later injected.

\begin{figure}[t]
\begin{center}
\includegraphics[width=\linewidth]{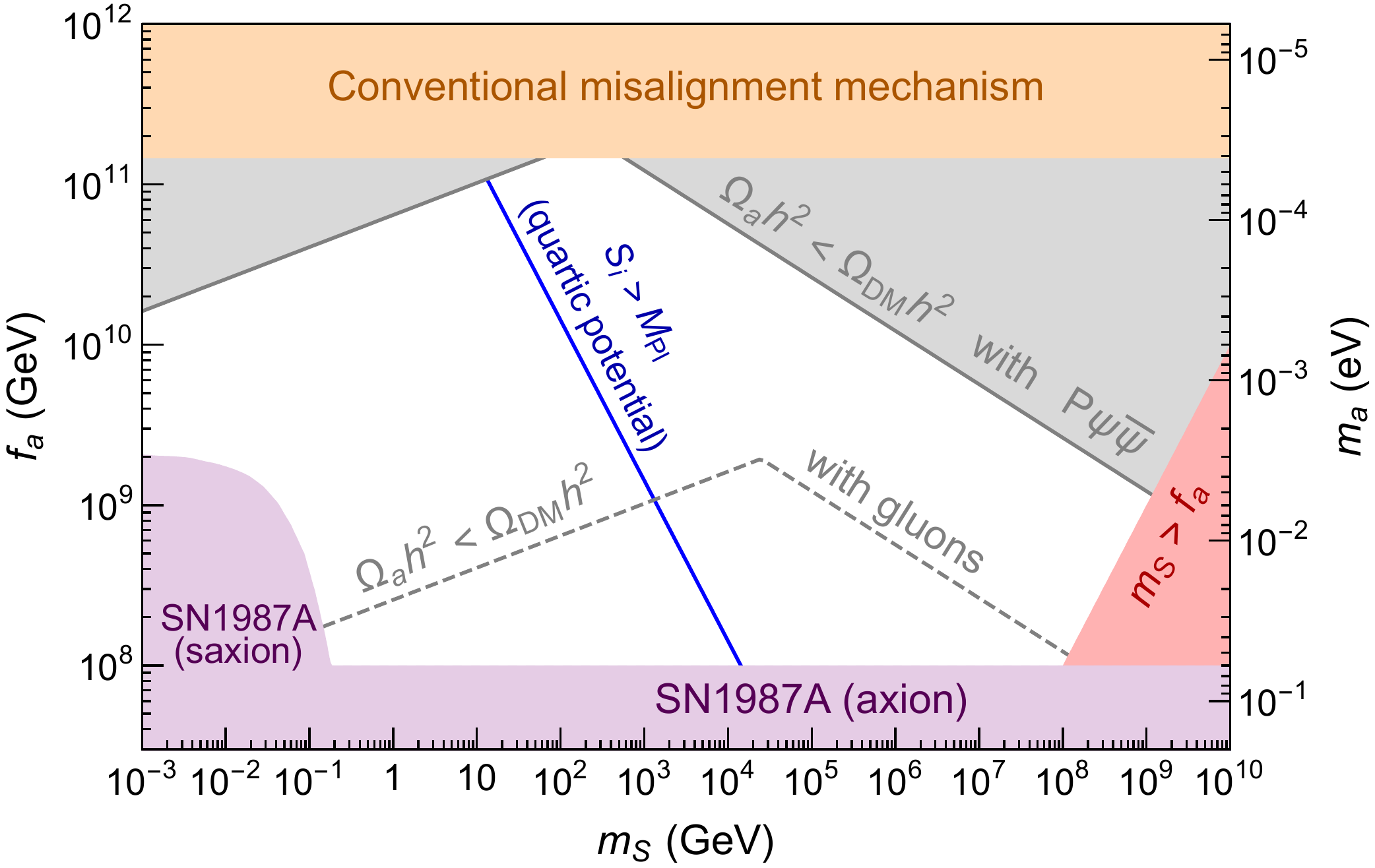}
\caption{
The parameter space of the QCD axion decay constant $f_a$ (or mass $m_a$) and the saxion vacuum mass $m_S$ compatible with the observed dark matter abundance. The blue line excludes high $m_S$ for the quartic potential. Applicable to both quartic and quadratic potentials, the gray region is ruled out for a maximal thermalization rate, while the constraint is the gray dashed line for thermalization via gluons only.
}
\label{fig:master}
\end{center}
\end{figure}

When $S \gg f_\phi$, the quartic term dominates and the energy density of the rotation redshifts as $R^{-4}$, $S \propto R^{-1}$, and $\dot\theta \propto R^{-1}$. When $ S \simeq f_\phi$, the quadratic term dominates and radial mode's energy density begins to redshift as $R^{-3}$ and $\dot\theta$ follows the usual scaling $R^{-3}$. Hence, a large initial $S$ slows down the redshift of $\dot\theta$. 

We assume that $P$ is thermalized to avoid overclosure from the radial mode. As shown in~\cite{Co:2019wyp}, even after thermalization, $P$ continues to rotate because it is energetically favorable to keep the charge asymmetry in the rotation rather than particle excitations. At thermalization, an elliptical trajectory becomes circular and the charge density $n_\theta$ stays conserved up to cosmic expansion. From charge conservation and the scaling of $P$, one finds that angular mode's energy density $\rho_\theta = n_\theta^2 / S^2$ decreases as $R^{-4}$ ($R^{-6}$) for $S \gg f_\phi$ ($S \simeq f_\phi$). The angular mode does not dominate the energy density since $\rho_\theta$ never redshifts slower than radiation.

The radius $S$ eventually settles to $f_\phi$, and the axion rapidly moves along the bottom of the potential in Eq.~(\ref{eq:quartic}).
The kinetic misalignment mechanism determines the axion abundance if $Y_\theta > Y_{\rm crit}$.

The available parameter space for QCD axion dark matter is summarized in Fig.~\ref{fig:master}, where various constraints are discussed in this and next sections. The red region violates unitarity of the saxion self interaction, while the purple region is excluded since the duration of the neutrino emission in a supernova core is altered by the emission of axions~\cite{Ellis:1987pk,Raffelt:1987yt,Turner:1987by,Mayle:1987as,Raffelt:2006cw,Chang:2018rso,Carenza:2019pxu} or saxions~\cite{Ishizuka:1989ts}. In the orange region, the conventional misalignment mechanism instead is operative since $Y_\theta < Y_{\rm crit}$ from Eqs.~(\ref{eq:YP_crit}) and (\ref{eq:Yv_axion}). The axion abundance is enhanced for larger $S_i$, but $S_i$ cannot exceed the Planck scale, giving an upper bound on $f_a$ based on Eqs.~(\ref{eq:Yv_axion}) and (\ref{eq:YP4})
\begin{equation}
\label{eq:Si_MPl}
f_a \lesssim 10^9 \GeV \ \epsilon^2 \left( \frac{\rm TeV}{m_S}  \right) ,
\end{equation}
corresponding to the blue line in Fig.~\ref{fig:master} for $\epsilon = 1$ and gets stronger if dilution due to entropy production is present. 

The energy density of the saxion must be depleted to avoid cosmological disaster, e.g.~excessive dark radiation from the decay to axions. Ensuring successful thermalization of $P$ further constrains the parameter space. Thermalization occurs when the temperature is still above the mass because $ m_S(T)/T$ is a constant and initially $m_S(T_{\rm osc}) \ll T_{\rm osc}$ based on Eq.~(\ref{eq:Tosc_Si}). Therefore, thermal dissipation plays an important role for thermalization~\cite{Yokoyama:2004pf,BasteroGil:2010pb,Drewes:2010pf,Mukaida:2012qn,Mukaida:2012bz}, rather than a  decay in the vacuum. The trajectory of $P$ becomes circular upon thermalization and we derive a consistency condition accordingly.
The dissipation rate by gluon scatterings is given by~\cite{Mukaida:2012qn}
\begin{equation}
\label{eq:Gamma_gluon}
\Gamma = b \frac{T^3}{f_{\rm eff}^2}, \hspace{0.5 in} b\simeq 10^{-5},
\end{equation}
where $f_{\rm eff} \equiv \sqrt{2}|P|$ after thermalization and is larger than $f_a$ at high temperatures.
Due to the scaling of Eq.~(\ref{eq:Gamma_gluon}), the thermalization rate relative to the Hubble scale is maximized when $f_{\rm eff}$ reaches $f_a$, implying a minimum temperature $T_{\rm min}$ at $f_{\rm eff} \simeq f_a$ and thus a maximum yield $Y_\theta^{\max} = m_S f_a^2/ s(T_{\rm min})$. Demanding that $Y_\theta^{\max}$ is sufficient for dark matter in Eq.~(\ref{eq:Yv_axion}) gives an upper bound
\begin{equation}
\label{eq:Ymax_DM}
f_a \lesssim 10^9 \GeV \,  \left( \frac{m_S}{{\rm TeV}} \right)^{ { \scalebox{1.01}{$\frac{1}{5}$} } } \left( \frac{b}{10^{-5}} \right)^{ \scalebox{1.01}{$\frac{3}{5}$} } ,
\end{equation}
which corresponds to the segment of the gray dashed line with a positive slope in Fig.~\ref{fig:master}. The constraint gets more stringent if entropy is injected after thermalization. To achieve $Y_\theta^{\max}$, $\epsilon$ is implicitly assumed to be large enough and the lower bound on $\epsilon$ follows from Eq.~(\ref{eq:Si_MPl}).

We next consider dissipation by a Yukawa coupling $y P \psi \bar{\psi}$. The dissipation rate is~\cite{Mukaida:2012qn}
\begin{equation}
\label{eq:Gamma_fermion}
\Gamma \simeq 0.1 \, y^2 \, T,
\end{equation}
where it is assumed that $m_\psi(T) \simeq y f_{\rm eff}(T) < T$. The maximal possible thermalization rate is then $0.1 \, T^3/ {f_{\rm eff}^2}$. A sufficient yield for dark matter gives the same bound as Eq.~(\ref{eq:Ymax_DM}) but with $b = 0.1$. This constraint is shown in Fig.~\ref{fig:master} by the gray region above the positively-sloped boundary. A wide range of $f_a \lesssim 10^{11}$ GeV is possible between such gray lines and the blue line from Eq.~(\ref{eq:Si_MPl}).

A sufficient amount of QCD axion dark matter requires that $m_S$ and hence the quartic coupling are small; namely the potential of $P$ is flat.
This is because a late start of the oscillation of $P$ enhances the charge to entropy ratio.

{\bf Supersymmetric models.}---%
The kinetic misalignment mechanism benefits from supersymmetry, where symmetry breaking fields naturally have flat potentials.

We consider the case where the saxion has a nearly quadratic potential with a typical mass $m_S$. This is the case for 1) a model with global symmetry breaking by dimensional transmutation due to the renormalization group running of the soft mass~\cite{Moxhay:1984am}, 
\begin{equation}
V = m_S^2 |P|^2 \left( {\rm ln}\frac{2 |P|^2}{f_\phi^2} -1  \right),
\end{equation}
2) a two-field model with soft masses,
\begin{equation}
W = X \left( P \bar{P}- V_P^2 \right), ~~V_{\rm soft} = m_P^2 |P|^2 + m_{\bar{P}}^2 |\bar{P}|^2,
\end{equation}
where $X$ is a chiral multiplet whose $F$-term fixes the global symmetry breaking fields $P$ and $\bar{P}$ along the moduli space $P \bar{P} = V_P^2$,
and 3) global symmetry breaking by quantum corrections in gauge mediation~\cite{ArkaniHamed:1998kj,Asaka:1998ns,Asaka:1998xa}.

For a nearly quadratic potential, the rotation of $P$ can occur in the same manner as the rotation of scalars in Affleck-Dine baryogenesis~\cite{Affleck:1984fy,Dine:1995kz}.
In the early universe $P$ may obtain a negative mass term by a Planck scale-suppressed coupling to the total energy density,
\begin{equation}
\label{eq:Hubble_mass}
V = - c_H H^2 |P|^2,
\end{equation}
where $H$ is the Hubble scale and $c_H$ is an $\mathcal{O}(1)$ constant.
For $H > m_S$, the saxion is driven to a large field value.

We consider explicit global symmetry breaking by a higher dimensional superpotential,
\begin{equation}
\label{eq:W}
W = \frac{P^{n+1}}{M^{n-2}}.
\end{equation}
The $F$-term potential from Eq.~(\ref{eq:W}) stabilizes the saxion $S \equiv \sqrt{2} |P|$  against the negative Hubble induced mass. The saxion tracks the minimum of the potential~\cite{Dine:1995kz,Harigaya:2015hha}
\begin{equation}
S(H) \simeq \left(H^2 M^{2n-4} \right)^\frac{1}{2n-2}.
\end{equation}
Once $H$ drops below $m_S$, the saxion begins to oscillate. Meanwhile, the supersymmetry breaking $A$-term potential associated with Eq.~(\ref{eq:W})
\begin{equation}
V \simeq (n+1) A \frac{P^{n+1}}{M^{n-2}} + {\rm h.c.},
\end{equation}
breaks the global symmetry explicitly, inducing the rotation of $P$. Here $A$ is of order the gravitino mass in gravity mediation.
According to Eq.~(\ref{eq:n_theta}), the asymmetry at the onset of the rotation is
\begin{equation}
n_\theta \simeq  A \ S(m_S)^2,
\end{equation}
if the initial phase is not accidentally aligned with the minimum.
At a large field value, the saxion mass tends to be dominated by the gravity mediated one, so $m_S = \mathcal{O}(A)$ and $n_\theta$ is of order $\rho_P/m_S$.
The charge density normalized by the saxion energy density remains constant despite the cosmic expansion.
(For $M = \mathcal{O}(M_{\rm Pl})$ and $A = \mathcal{O}({\rm TeV})$, a shift to the CP violating phase of the strong interaction from the explicit PQ symmetry breaking is smaller than the experimental upper bound if $n > 7 - 9$ for $f_a = 10^9-10^{12}$ GeV.)

Due to the large initial field value, the saxion tends to dominate the energy density of the Universe, which we assume hereafter.
Regardless, $P$ has to be thermalized eventually. After thermalization completes at the temperature $T_{\rm th}$, $P$ rotates with a vanishing ellipticity and with the total charge $n_\theta R^3  = S^2 \dot\theta R^3$ conserved. The charge conservation implies that $\dot\theta$ stays constant for $|P| \gg f_\phi$ before following the usual $R^{-3}$ scaling when $|P| \simeq f_\phi$. Thermalization transfers the energy of the radial motion of $P$ into radiation. The remaining energy is associated with a circular motion, $\rho_\theta$. The final yield is
\begin{equation}
\label{eq:YP2}
Y_\theta  = \frac{n_\theta}{s} = \epsilon \frac{3 T_{\rm th}}{4 m_S} ,  \hspace{0.3in}
\epsilon  \equiv \frac{n_\theta}{\frac{\rho_P}{m_S} - n_\theta} \simeq \frac{A}{m_S} ,
\end{equation}
and $\epsilon \lesssim 1$ measures the amount of angular rotations relative to radial oscillations.

After thermalization, the equation of motion fixes $\dot{\theta} = m_S$, with which one can easily show by conservation of energy and $U(1)$ charge that $|P|$ does not immediately drop to $f_\phi$ as usual thermalization does for a scalar without a $U(1)$ charge. Instead, $|P|$ redshifts by the cosmic expansion. The energy density of the circular rotation decreases as $R^{-3}$ ($R^{-6}$) for $|P| \gg f_\phi$ ($|P| \simeq f_\phi$). Right after thermalization, the universe is still dominated by the circular rotation, but after the $R^{-6}$ scaling begins, the Universe is eventually dominated by the thermal bath created by the aforementioned thermalization process.

We focus on the QCD axion and discuss whether sufficient axion dark matter can be produced. From Eqs.~(\ref{eq:Yv_axion}) and (\ref{eq:YP2}), the thermalization temperature needed to obtain the observed dark matter abundance is
\begin{equation}
\label{eq:TthDM}
T_{\rm th} \simeq  50 \ \frac{m_S}{\epsilon} \left( \frac{f_a}{10^9~{\rm GeV}} \right).
\end{equation}
Since $T_{\rm th} \gg m_S$, thermal dissipation is necessary. 

Relevant for the thermalization rate, the effective decay constant just after thermalization can be obtained by energy and charge conservation, and reads
\begin{equation}
f_{\rm eff}^2 = 2 |P|^2  \simeq  \max \left[ f_a^2, \ \epsilon \frac{\pi^2 g_*}{60} \frac{T_{\rm th}^4}{m_S^2} \right],
\end{equation}
We begin with the scenario where thermalization occurs via scattering with gluons, with the rate given in Eq.~(\ref{eq:Gamma_gluon}). Thermalization completes when $\Gamma = 3H$. From the dependence of $\Gamma$ on $T$, one can see that thermalization is possible only if $f_{\rm eff} > f_a$, for which
\begin{equation}
T_{\rm th} =10^3~{\rm GeV} \left( \frac{b}{10^{-5}}\right)^{ \scalebox{1.01}{$\frac{1}{3}$} } \left( \frac{m_S}{\rm TeV}\right)^{ \scalebox{1.01}{$\frac{2}{3}$} }.
\end{equation}
To obtain the dark matter abundance, this actual thermalization temperature has to be above or equal to that in Eq.~(\ref{eq:TthDM}). (In the former case, the correct abundance can be obtained without matter domination by $P$ or extra dilution.) This leads to an upper bound on $f_a$
\begin{equation}
\label{eq:famax_DM}
 f_a \lesssim 10^9 \GeV \ \epsilon^{2/3} \left( \frac{10^5 \GeV}{m_S} \right)^{ \scalebox{1.01}{$\frac{1}{3}$} } \left( \frac{b}{10^{-5}} \right)^{ \scalebox{1.01}{$\frac{1}{3}$} } .
\end{equation}
This constraint with $\epsilon =1$ is shown in Fig.~\ref{fig:master} by the negatively-sloped gray dashed line. The bound in Eq.~(\ref{eq:Ymax_DM}) also applies to the nearly quadratic potential, shown by the positively-sloped gray dashed line. The decay constant is predicted to be below $\mathcal{O}(10^9)$ GeV. Remarkably, the required saxion mass comes out consistent with TeV-PeV scale supersymmetry.

For dissipation by $y P \psi \bar{\psi}$ scattering, the rate is given in Eq.~(\ref{eq:Gamma_fermion}) and we again obtain Eqs.~(\ref{eq:Ymax_DM}) and (\ref{eq:famax_DM}) but with $b = 0.1$. This constraint is shown in Fig.~\ref{fig:master} by the gray shaded region. A wider range of $f_a \lesssim 10^{11}$ GeV becomes possible for fermion $\psi$ scatterings (shown by the negatively-sloped gray boundary) compared to gluon scatterings. In supersymmetric models, larger values of $m_S$ become viable compared to the case of the quartic potential (blue line). 

{\bf Discussion.}---%
We presented the kinetic misalignment mechanism, where the dark matter abundance of a generic axion-like particle is determined by the initial field velocity, as opposed to the conventionally assumed initial misalignment. 
We then showed that this can yield QCD axion dark matter for any $f_a$ below $1.5 \times 10^{11}$ GeV, down to the minimum value allowed by supernovae constraints, $f_a \sim 10^8$ GeV. We studied in detail the full cosmological evolution of the PQ field in both quartic and quadratic potentials, with the results in Figure \ref{fig:master} showing that kinetic misalignment is successful over a wide range of parameters.
Besides signals in axion searches in the mass range ${\cal O}(0.1-100)$ meV, kinetic misalignment can provide a unified origin of dark matter and the cosmological excess of matter over antimatter~\cite{Co:2019wyp}.

Here we draw a connection of kinetic misalignment by higher dimensional operators with parametric resonance production we discussed in~\cite{Co:2017mop}. In both mechanisms, we assume a large initial field value of the symmetry breaking field in the early universe. In~\cite{Co:2017mop}, it is assumed that the global symmetry is a good symmetry and the saxion simply oscillates through the origin. Then axion dark matter is produced by parametric resonance. In this Letter, we assume alternatively that the global symmetry is explicitly broken and a rotation is induced. The kinetic misalignment mechanism can hence be understood as complementary to the production by parametric resonance. Explicit PQ breaking might also lead to a signal of a nonzero neutron electric dipole moment.

Throughout the Letter, we assumed the field rotation remains coherent. If the radial direction of the global symmetry breaking field has a potential flatter than quadratic, an instability develops and the coherent rotation fragments into inhomogeneous configurations~\cite{Kusenko:1997si,Enqvist:1997si,Enqvist:1998en,Kasuya:1999wu}. In our case, it is unclear how the inhomogeneity evolves since the symmetry is spontaneously broken in the vacuum. In fact, the standard Q-ball solution, where the rotating field has a vanishing field value at infinity, does not have a finite energy density in this case. It will be interesting to investigate the fate of the inhomogeneity and its impact on the axion abundance.

{\bf Acknowledgment.}---%
The work was supported in part by the DoE Early Career Grant DE-SC0019225 (R.C.), the DoE grants DE-AC02-05CH11231 (L.H.) and DE-SC0009988 (K.H.), the NSF grant NSF-1638509 (L.H.), as well as the Raymond and Beverly Sackler Foundation Fund (K.H.).

\bibliography{KMM} 

\clearpage
\maketitle
\onecolumngrid
\begin{center}
\textbf{\large Axion Kinetic Misalignment Mechanism} \\ 
\vspace{0.05in}
{ \it \large Supplemental Material}\\ 
\vspace{0.05in}
{Raymond T. Co, Keisuke Harigaya, and Lawrence J. Hall}
\end{center}
\onecolumngrid
\setcounter{equation}{0}
\setcounter{figure}{0}
\setcounter{table}{0}
\setcounter{section}{0}
\setcounter{page}{1}
\makeatletter
\renewcommand{\theequation}{S\arabic{equation}}
\renewcommand{\thefigure}{S\arabic{figure}}

This Supplemental Material is organized as follows. In Sec.~\ref{sec:CMM}, we review the axion field dynamics in the conventional misalignment mechanism setting. In Sec.~\ref{sec:KMM}, we derive the formulae for the axion dark matter abundance from the rotating Peccei-Quinn symmetry breaking field both analytically and numerically. 

\section{Review of the Conventional Misalignment Mechanism}
\label{sec:CMM}

The potential the potential energy of the axion $V(\phi)$ is periodic in the axion field space and takes the form
\begin{equation}
V(\phi) = m_\phi^2 f_\phi^2 \left( 1 - \cos \left( \frac{\phi}{f_\phi} \right) \right) .
\end{equation}
In this section, we assume a constant mass $m_\phi$ for illustration purposes. A dimensionless variable conveniently defined by $\theta \equiv \phi / f_\phi$ is called the axion misalignment angle and measures the angular displacement from the minimum of the cosine potential. The axion field dynamics in an expanding Universe obeys the equation of motion given by
\begin{equation}
\label{eq:eom}
\ddot\phi + 3 H \dot\phi + V'(\phi) = 0 ,
\end{equation}
where the dot denotes coordinate-time derivative and it is assumed that the axion field is coherent and thus the spatial gradient vanishes. The conventional misalignment mechanism assumes the initial condition at $t = t_i \ll m_\phi^{-1}$
\begin{align}
\theta (t_i) &= \theta_i \\
\dot\theta (t_i) &= 0 .
\end{align}

In a radiation-dominated universe, the Hubble expantion rate is given by
\begin{align}
\label{eq:H_t}
H = \frac{1}{2t} ,
\end{align}
which leads to the solution
\begin{align}
\theta(t) & = 2^{1 / 4} \Gamma\left(\frac{5}{4}\right) \frac{J_{1 / 4}(m_\phi t)}{(m_\phi t)^{1 / 4} } \times \theta_i \\ 
& \approx \theta_i \times \begin{cases}
\ \ \ \ \ \ \ \ \ \ \ \ 1 \qquad  & \text{for} \quad m_\phi t \ll 1 \\
  2^{1 / 4} \Gamma\left(\frac{5}{4}\right) \sqrt{\frac{2}{\pi}}   \left(\frac{1}{m_\phi t} \right)^{3 / 4}  \cos \left(m_\phi t -\frac{3}{8} \pi\right) \qquad  & \text{for} \quad m_\phi t \gg 1 
  \end{cases} .
\end{align}
This makes it evident that the axion field is frozen at early times due to the Hubble friction term in Eq.~(\ref{eq:eom}), which is analogous to an overdamped harmonic oscillator. The oscillation occurs when Hubble friction drops below the mass term, {\it i.e.}~$3 H(T_*) \simeq m_\phi$, at the temperature we call $T_*$. After the onset of oscillations, the amplitude of $\theta(t)$ scales as $t^{-3/4} \propto R^{-3/2}$ due to redshift, where $R$ is the scale factor of the universe. The energy density of the axion
\begin{align}
\rho_\phi & = \frac{1}{2} \dot\phi^2 + V(\phi)  \\
& \approx f_a^2 \left( \frac{1}{2} \dot\theta^2 + \frac{1}{2} m_\phi^2 \theta^2 \right) \qquad   \text{for} \quad \theta \ll 1
\end{align}
scales as that of matter after $T_*$. It is common to compute the redshift-invariant quantity 
\begin{equation}
\frac{\rho_\phi}{s} = \frac{\rho_\phi(t_i) }{s(T_*)} ,
\end{equation}
where $s(T) = \frac{2\pi^2}{45} g_* T^3$ is the entropy density, $g_*$ is the effective number of degrees of freedom in the thermal bath. The observed dark matter abundance today is 
\begin{equation}
\label{eq:rho_DM}
\frac{\rho_{\rm DM}}{s}  = 0.44 \, {\rm eV} \, \simeq T_e,
\end{equation}
and $T_e$ is the temperature at matter-radiation equality.

The derivation is generalized as follows for the case where the axion mass $m_\phi$ is time-dependent, as is the case for the QCD axion when the temperature is above the QCD confinement scale $\Lambda_{\rm QCD}$. If the mass changes adiabatically, {\it i.e.}~$\dot m_\phi \ll m_\phi^2$, then the final abundance is given by
\begin{equation}
\label{eq:rho_mT}
\frac{\rho_\phi}{s} = m_\phi(T=0) \frac{n_\phi (T_*)}{s(T_*)},
\end{equation}
with $n_\phi (T_*)$ the number density of $\phi$ at $T_*$.

\section{Detailed Analyses of the Kinetic Misalignment Mechanism}
\label{sec:KMM}

The kinetic misalignment mechanism operates when the initial angular speed $\dot\theta_i$ is nonzero and sufficiently large. The origin of this alternative initial condition is well motivated and discussed in the main Letter. Assuming $\dot\theta_i \neq 0$, we give the critical value of $\dot\theta_i$ such that the axion dynamics is altered from the conventional case discussed in Sec.~\ref{sec:CMM} and derive the expression for the axion abundance. We will start with a simplified analytic treatment as done in the main Letter, perform a rigorous numerical computation, and finally provide an improved analytic understanding based on the numerical result.

If the axion kinetic energy $K = \dot\theta^2 f_\phi^2/2$ is larger than its potential energy $V(\phi)$ at the conventional oscillation temperature $T_*$, the axion simply overcomes the potential barrier and the misalignment angle continues to change at the rate $\dot\theta (T)$. This evolution ceases when the kinetic energy $K$ redshifts to below the potential barrier $V(\phi) = 2 m_\phi^2(T) f_\phi^2$ at the temperature we call $T'$,
\begin{equation}
\label{eq:dtheta_mphi}
\dot\theta (T') = 2 m_\phi(T'), 
\end{equation}
after which the axion is trapped due to the potential barrier and oscillates around the minimum. The onset of oscillation is delayed if this trapping happens after the conventional oscillation temperature, $T' < T_*$. Equivalently, kinetic misalignment is at play when 
\begin{equation}
\label{eq:mphi_Hubble}
m_\phi(T') \geq 3 H(T') = \sqrt{ \frac{\pi^2}{10} g_*} \frac{T'^2}{M_{\rm Pl}}.
\end{equation}

Using Eqs.~(\ref{eq:rho_mT}) and (\ref{eq:dtheta_mphi}), the abundance is estimated by
\begin{align}
\frac{\rho_\phi}{s} & = m_\phi(T=0) \frac{V_\phi / m_\phi(T')}{s(T')} = m_\phi(T=0) Y_\theta,
\label{eq:rho_s_dtheta}
\end{align}
where we define
\begin{equation}
\label{eq:Ytheta}
Y_\theta \equiv \frac{\dot\theta f_\phi^2}{s} ,
\end{equation}
a quantity that is associated with the $U(1)$ charge density $n_\theta = \dot\theta f_\phi^2$, {\it e.g.}~the Peccei-Quinn charge density for the QCD axion. Even though $Y_\theta$ is a redshift-invariant quantity, the evaluation of $Y_\theta$ from the initial condition is model-dependent and is presented in the main Letter.

The analytic estimate thus far is not perfectly precise because $\rho_\phi$ is assumed to scale as $R^{-3}$ as soon as the oscillation starts at $T'$. Since the oscillation starts near the hilltop of the cosine potential, it is known that anharmonicity near the hilltop enhances the axion abundance. This can simply be understood as a further delay of the oscillation because the potential gradient vanishes around the hilltop. This nonlinear effect calls for a numerical analysis, which we perform as follows. We solve the equation of motion given in Eq.~(\ref{eq:eom}), where we assume large values of $\dot\theta_i$ that safely satisfy Eq.~(\ref{eq:mphi_Hubble}) and use the QCD axion mass scaling 
\begin{equation}
\label{eq:maT}
m_a (T) = 6 \, {\rm meV} \, \left( \frac{10^9 \, {\rm GeV}}{f_a} \right) \times
\begin{cases}
\ \ \ \ 1 \qquad  & \text{for} \quad T \leq \Lambda_{\rm QCD} \\
\left( \frac{\Lambda_{\rm QCD}}{T}\right)^n  \qquad  & \text{for} \quad T \geq \Lambda_{\rm QCD}
\end{cases} ,
\end{equation}
with $n=4$ from the dilute instanton gas approximation and $\Lambda_{\rm QCD} = 150$ MeV. Our numerical result confirms the analytic scaling observed in Eq.~(\ref{eq:rho_s_dtheta}), whereas the overall normalization $C$ of the axion abundance defined by
\begin{equation}
\label{eq:rho_s_dtheta_num}
\frac{\rho_\phi}{s} = C m_\phi(T=0) Y_\theta 
\end{equation} 
is numerically determined as $C \simeq 2$ in the following way.

We solve the equation of motion
\begin{equation}
\label{eq:eom_num}
\ddot\theta + 3 \frac{1}{2t} \dot\theta = - m_a^2(t) \sin (\theta) ,
\end{equation}
where we have assumed a radiation-dominated universe and used the Hubble parameter in Eq.~(\ref{eq:H_t}) to relate time and temperature for the axion mass given in Eq.~(\ref{eq:maT}). Throughout the computation, we fix $g_*(\Lambda_{\rm QCD}) \simeq 26$ even though the final result is insensitive to this. The initial condition for $\dot\theta$ is computed using the value of $Y_\theta$ defined in Eq.~(\ref{eq:Ytheta}) at a time around one order of magnitude before the oscillation time $\dot\theta \simeq m_a$. The initial condition for $\theta$ is irrelevant because the final abundance is solely determined by $\dot\theta$. We evaluate the axion yield $Y_a \equiv n_a/s$ with $ n_a = \rho_a(t_f) / m_a(t_f)$ at a final time $t_f$ when the transient behavior is sufficiently damped. With this final conserved quantity $Y_a$, we compare the zero-temperature axion energy density numerically obtained $\rho_a = m_a(T = 0) Y_a$ with the formula in Eq.~(\ref{eq:rho_s_dtheta_num}) to determine the value of $C$.

\begin{figure}
	\includegraphics[width=\linewidth]{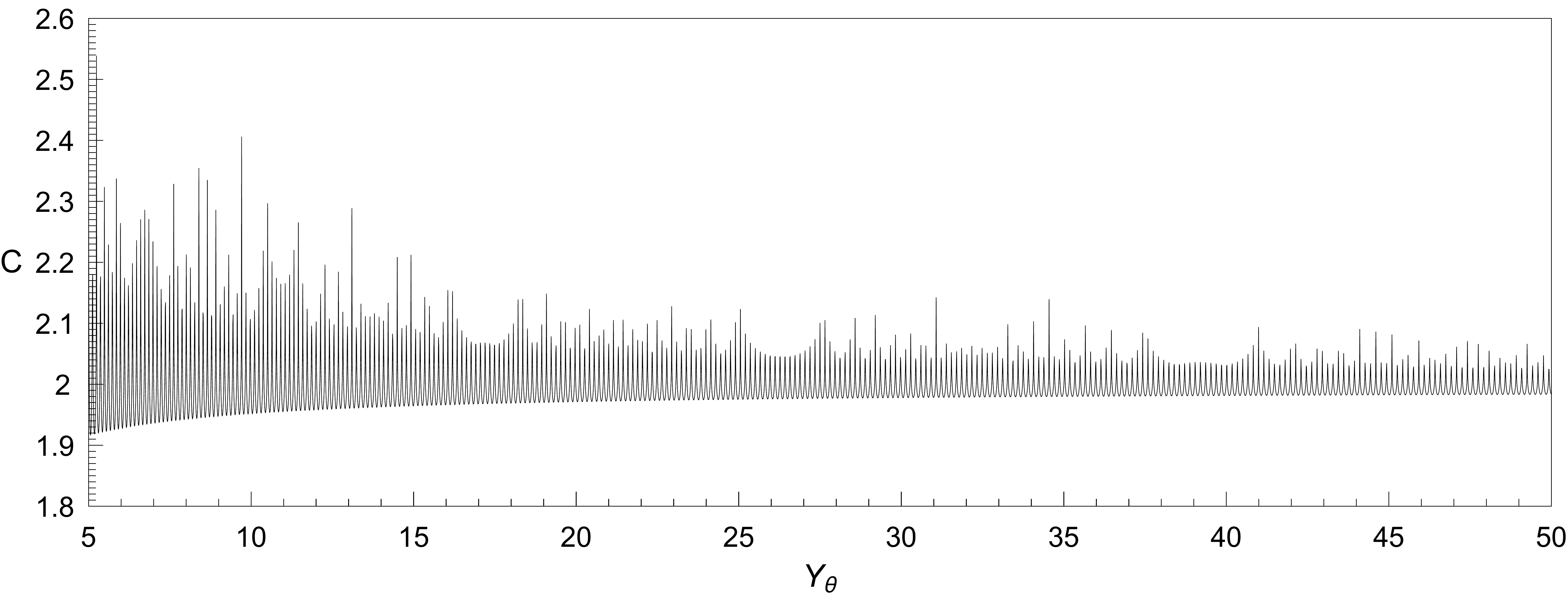}
	\includegraphics[width=\linewidth]{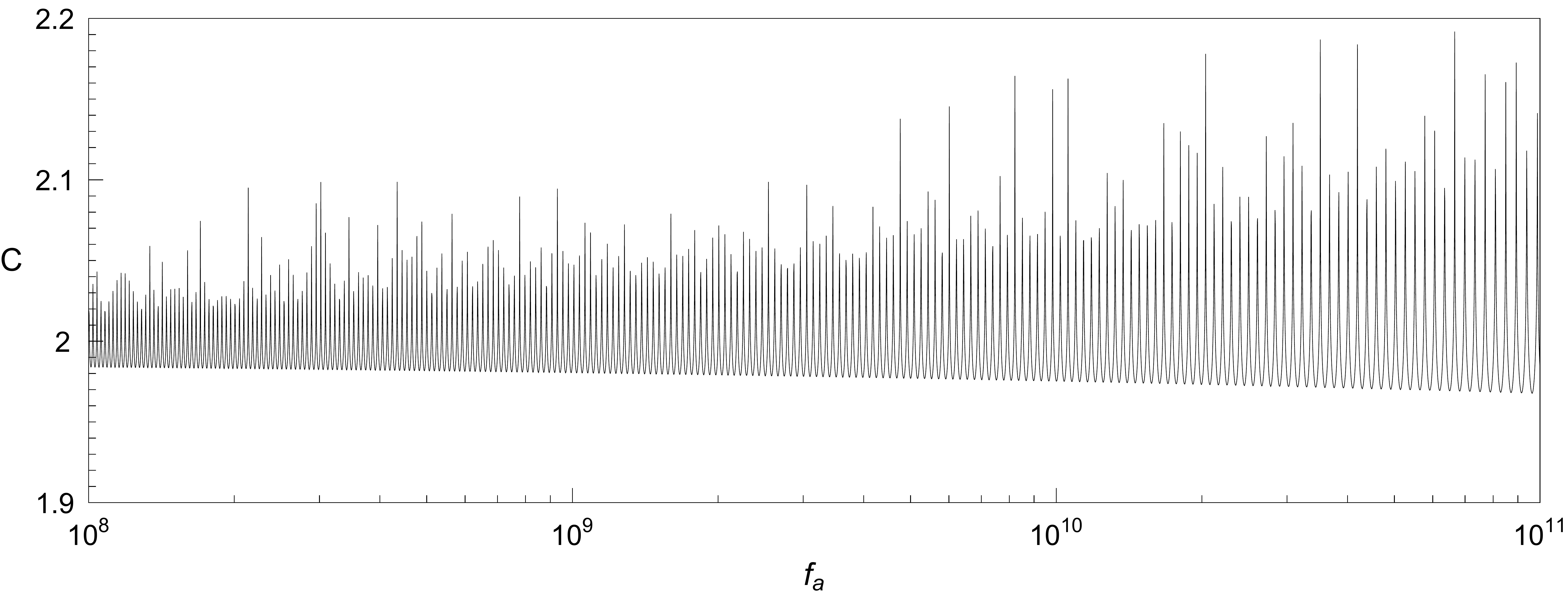}
	\caption{Values of $C$ determined by numerical computations. In the upper panel, we fix $f_a = 10^9$ GeV and vary $Y_\theta$. In the lower panel, we vary $f_a$ while fixing $\dot\theta$ to the value such that $f_a = 10^9$ reproduces the observed dark matter abundance.}
	\label{fig:C}
\end{figure}

In Fig.~\ref{fig:C}, we show the values of $C$ for different values of $Y_\theta$ and $f_a$. In the upper panel, we fix $f_a$ and vary $Y_\theta$ from 5 to 50; note that the observed dark matter abundance is reproduced for $Y_\theta \simeq T_e / C m_a \simeq 36.7$. In the lower panel, we vary $f_a$ from $10^8$ GeV to $10^{11}$ GeV while fixing the initial condition of $\dot\theta$ such that the observed dark matter abundance is obtained for $f_a = 10^9$ GeV using $C = 2$. Fig.~\ref{fig:C} shows that average value centers around $C = 2$ while the nonlinear effects lead to a range of $1.92 \lesssim C \lesssim 2.54$. We note however that if we restrict the parameter space to the contour for the observed dark matter abundance by allowing $f_a$ to vary accordingly, the value of $C = 2$ is a good approximation with an accuracy better than 10\% for $f_a \lesssim 2 \times 10^{10}$ GeV and the accuracy worsens to 20-30\% around $f_a = 10^{11}$ GeV.

\begin{figure}
	\includegraphics[width=0.49\linewidth]{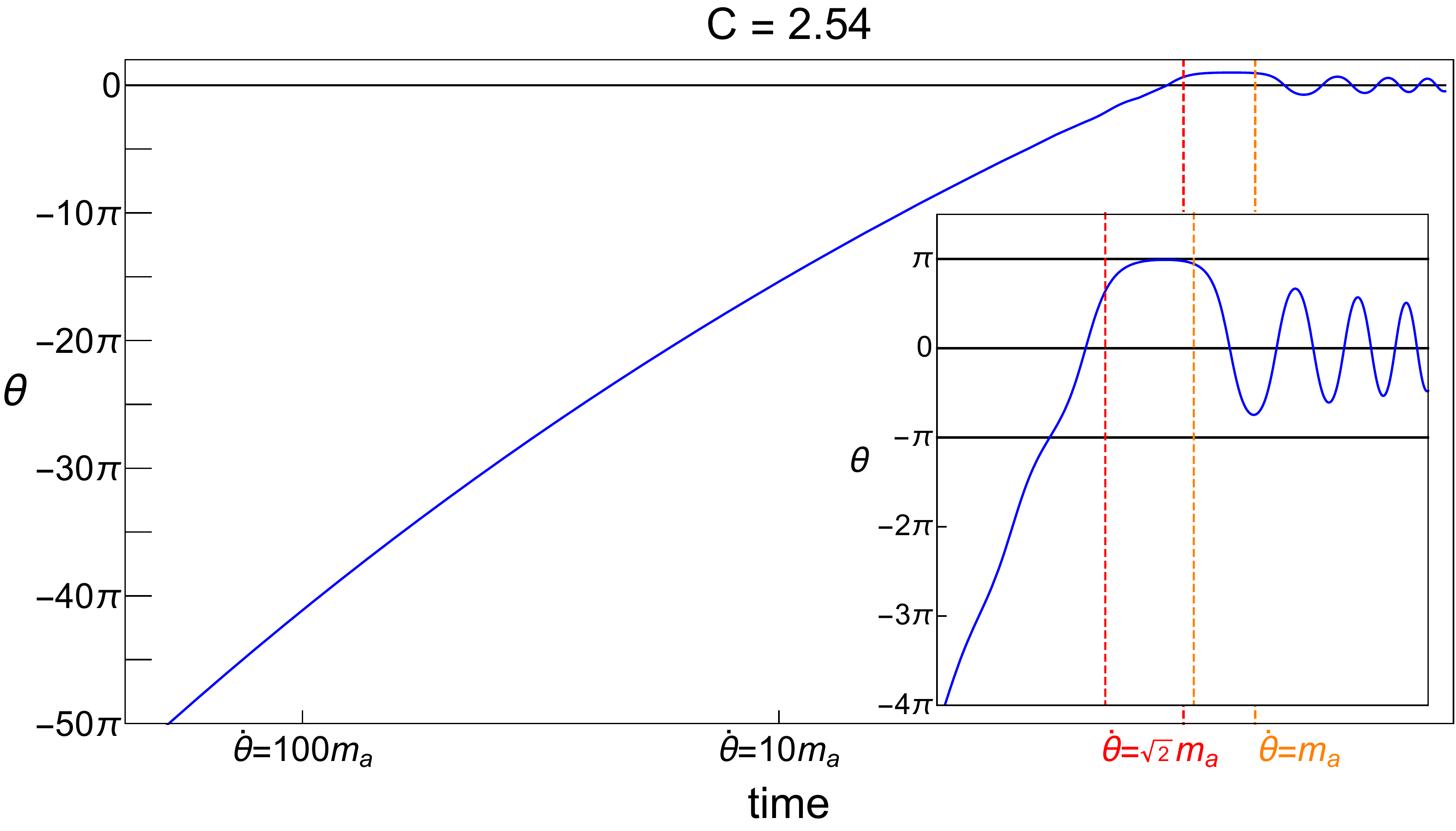}
	\includegraphics[width=0.49\linewidth]{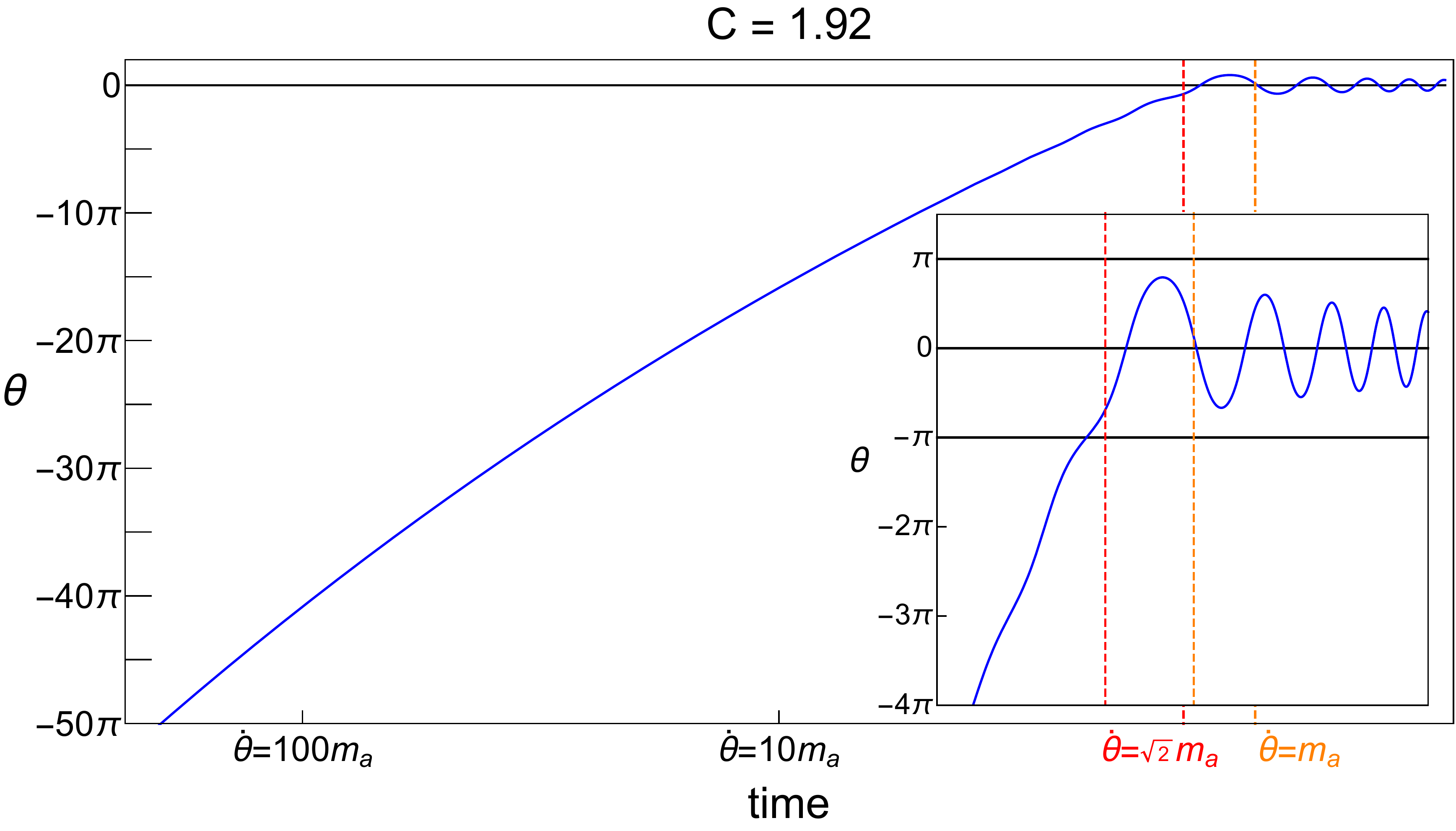}
	\caption{The evolution of the misalignment angle $\theta$ for the maximum (left panel) and minimum (right panel) of the values of $C$ obtained in the numerical scan in the upper panel of Fig.~\ref{fig:C}.}
	\label{fig:theta_max_min}
\end{figure}

In Fig.~\ref{fig:theta_max_min}, we study the cases where a large or small value of $C$ arises. In the left (right) panel, we use the benchmark point that corresponds to the maximum (minimum) value of $C$ obtained from the scan for the upper panel of Fig.~\ref{fig:C}, both of which occur at low $Y_\theta \simeq 5$. As seen in the left panel, larger values of $C$ result because the field transitions from the rotation to the oscillation in close proximity to the hilltop of the potential, where the anharmonicity effect is strongest. Smaller values of $C$, on the other hand, arise when such a transition occurs on the hillside instead as demonstrated in the right panel.

The transition from the rotation to the oscillation when $\dot\theta \simeq \sqrt{2} m_a$ can also be understood by a more detailed analytic treatment. When the kinetic energy estimate $K_{\rm est} = \dot\theta^2 f_a^2/2$ obtained using a simple $\dot\theta$ scaling is comparable to the potential energy $V(a)$, the true kinetic energy $K$ in fact fluctuates because the axion field is running up and down the potential barriers $V_{\rm bar} = 2 m_a^2 f_a^2$. The transition occurs when the upward-fluctuated kinetic energy $K_{\rm est} + V_{\rm bar} / 2$ becomes insufficient in overcoming the potential barrier $V_{\rm bar}$. This leads to the condition $\dot \theta \simeq \sqrt{2} m_a$. This transition condition, however, would predict analytically that $C \simeq \sqrt{2}$ and thus still does not fully explain the numerically-determined value of $C \simeq 2$. We attribute the remaining discrepancy to the enhancement from anharmonicity effects the analytic estimate does not capture.

Before concluding this section, we now revisit the condition for kinetic misalignment. With Eqs.~(\ref{eq:rho_DM}), (\ref{eq:dtheta_mphi}), and (\ref{eq:rho_s_dtheta_num}), the condition for the kinetic misalignment mechanism in Eq.~(\ref{eq:mphi_Hubble}) can be rewritten as
\begin{equation}
m_\phi(T=0) f_\phi^2 \leq \frac{2}{9 C} \sqrt{\frac{\pi^2 g_*}{10}}  T_e T_*  M_{\rm Pl}.
\end{equation}
When this condition is applied to the QCD axion $a$, kinetic misalignment is effective when the yield is larger than the critical value
\begin{equation}
\label{eq:Y_crit_full}
Y_{\rm crit} = \frac{135 C}{2 \pi^2 g_*} \left( \frac{3 f_a^{12}}{m_a(T=0) M_{\rm Pl}^7 \Lambda_{\rm QCD}^4 } \right)^{ \scalebox{1.01}{$\frac{1}{6}$} } ,
\end{equation}
which is obtained using Eq.~(\ref{eq:rho_s_dtheta_num}) and $\dot\theta(T_*) > 2 m_a (T_*)$ so that the kinetic energy dominates over the potential energy at the usual oscillation temperature
\begin{equation}
T_* = \left( \sqrt{\frac{10}{g_*}} \frac{m_a(T=0) M_{\rm Pl} \Lambda_{\rm QCD}^4}{\pi} \right)^6 .
\end{equation}
Kinetic misalignment can only account for the dark matter abundance when
\begin{equation}
f_a \lesssim 1.5 \times 10^{11} \, {\rm GeV}  \left( \frac{2}{C} \right)^{ \scalebox{1.01}{$\frac{12}{7}$} } \left( \frac{g_* (\Lambda_{\rm QCD})}{30} \right)^{ \scalebox{1.01}{$\frac{5}{14}$} } \left( \frac{\Lambda_{\rm QCD}}{150 \, {\rm MeV}} \right)^{ \scalebox{1.01}{$\frac{4}{7}$} } 
\end{equation}
because otherwise the yield required by Eqs.~(\ref{eq:rho_DM}) and (\ref{eq:rho_s_dtheta_num}) is smaller $Y_{\rm crit}$ in Eq.~(\ref{eq:Y_crit_full}).

\end{document}